\documentclass[]{article}

\usepackage[utf8]{inputenc}

\usepackage{fullpage}

\usepackage{amsmath}
\usepackage{amssymb}
\usepackage{amsthm}
\usepackage{latexsym}
\usepackage[font=small,labelfont=bf,labelsep=space]{caption}
\usepackage{float}
\newtheorem{definition}{Definizione}

\usepackage{boxedminipage}

\usepackage{listings}

\usepackage{minitoc}

\usepackage{ifpdf}

\ifpdf
\usepackage[pdftex]{graphicx}
\else
\usepackage{graphicx}
\fi
\title{Line graphs e contrazioni: un approccio rigoroso alla space syntax}
\author{Roberto D'Autilia,\\
www.formulas.it: Laboratorio di Matematica,\\
Dipartimento di Architettura,\\
Universit\`a degli Studi Roma Tre\\
roberto.dautilia@uniroma3.it}
\date{2015-03-14}

\begin{document}
	\lstset{
	language=Mathematica,                             
	basicstyle=\ttfamily,                   
	columns=flexible,                       
	}
\ifpdf
\DeclareGraphicsExtensions{.pdf, .jpg, .tif}
\else
\DeclareGraphicsExtensions{.eps, .jpg}
\fi

\maketitle

\begin{abstract}
I metodi della {\sl space syntax} sono stati oggetto di una vasta discussione e di diverse proposte di modifica degli algoritmi per la costruzione delle linee assiali.
La {\sl space syntax} è rappresentabile in termini di grafi definiti sugli {\sl edge} di un grafo primario ({\sl line graph}).
Per mezzo degli algoritmi di {\sl line graph}, un sistema di etichette definite sugli {\sl edge} del grafo primario si trasforma in un sistema di etichette sui vertici del {\sl line graph}.
Un algoritmo di contrazione di {\sl edge} adiacenti con la stessa etichetta permette di costruire un grafo più generale di quelli generati con i metodi della {\sl space syntax}.
Per mezzo delle funzioni implementate in Mathematica è quindi possibile ridefinire la {\sl space syntax} su un sistema qualsiasi di variabili (etichette) e riprodurre i risultati delle linee assiali come caso particolare. Come esempio viene discussa una possibile applicazione acustica del metodo.
\end{abstract}

\section{Introduzione}
La teoria della {\sl space syntax} raccoglie alcuni metodi formali per studiare l'accessibilit\`a e altre propriet\`a degli spazi urbani.
Partendo dal lavoro di Hillier \cite{HillierSLS} la {\sl space syntax} \`e stata sviluppata e applicata a numerosi casi di studio e ha stimolato la produzione di {\sl software} che ne implementano i metodi.

Le variabili sulle quali si fonda la teoria sono le isoviste e le corrispondenti linee assiali che interagiscono per mezzo della loro intersezione \cite{RePEc:pio:envirb:v:34:y:2007:i:3:p:539-555}.
Una linea assiale identifica uno spazio aperto.
Quando due linee assiali si intersecano, allora si ha un'interazione tra i due spazi corrispondenti che d\`a luogo ad una connessione diretta e quindi alla possibilità di passare da uno spazio all'altro.

Questo modello dello spazio urbano ha fornito una grande quantit\`a di risultati e un punto di vista originale sulla struttura e sulle funzioni della citt\`a \cite{taus}\cite{RePEc:pio:envirb:v:34:y:2007:i:3:p:539-555}\cite{PhysRevE.73.066107}, ma ha anche dato vita a un dibattito sulla sua generalit\`a e su alcune potenziali incongruenze \cite{Ratti2004}\cite{RePEc:pio:envirb:v:31:y:2004:i:4:p:501-511}\cite{RePEc:pio:envirb:v:31:y:2004:i:4:p:513-516}.
Alcuni autori hanno analizzato il possibile conflitto tra la rappresentazione delle caratteristiche topologiche e geometriche dello spazio urbano, evidenziando una apparente mancanza di robustezza del metodo rispetto a piccole modifiche della geometria della citt\`a \cite{Ratti2004}\cite{RePEc:pio:envirb:v:31:y:2004:i:4:p:501-511}.

In ambito fisico-matematico \`e stato invece mostrato come i modelli della {\sl space syntax} possono essere riformulati in termini di teoria dei grafi \cite{BV_Buch}, associando un nodo ad ogni linea assiale e connettendo due nodi con un {\sl edge} se le corrispondenti linee si intersecano.
Inoltre \`e stato anche mostrato che questa analisi dello spazio urbano \`e legata all'esplorazione casuale, cio\`e alle misure stazionarie di un {\sl random walk} su un grafo \cite{hillierparadigm1999}, e sono stati realizzati diversi strumenti numerici \cite{VaroudisX} che identificano questi grafi a partire dagli spazi aperti, dalle loro linee assiali e dalle loro intersezioni.

Se definiamo un grafo primario $G_p$ su una mappa urbana come il grafo che localizza i nodi sugli incroci stradali e identifica gli {\sl edge} con strade e piazze, i grafi assiali sono legati, in un modo che vedremo in dettaglio nei prossimi paragrafi, al {\sl line graph} di $G_p$, cio\`e al grafo sugli {\sl edge} tale che due nodi sono connessi se i corrispondenti {\sl edge} del grafo primario sono adiacenti \cite{Diestel2005}.
Sembra quindi naturale cercare un procedimento rigoroso per ottenere i risultati della {\sl space syntax} partendo solamente dai grafi primari e dai corrispondenti {\sl line graph}.

In questo ambito sono state fatte diverse proposte con lo scopo principale di risolvere il problema di trovare una continuit\`a strutturale tra spazi adiacenti che potrebbero non essere connessi da un'unica linea assiale, proposte che vanno dalla misura degli angoli di inclinazione tra le linee assiali \cite{RePEc:pio:envirb:v:34:y:2007:i:3:p:539-555} all'introduzione di propriet\`a definite sugli {\sl edge}, come per esempio i nomi delle vie \cite{Porta2006}.

Per risolvere questo problema abbiamo introdotto e implementato in Mathematica un metodo per ottenere i risultati della {\sl space syntax} direttamente dai grafi primari, sfruttando le funzioni per la costruzione di {\sl line graph} e la contrazione di grafi.
Gli ingredienti del modello sono i grafi primari, facilmente identificabili direttamente o con algoritmi di processamento di immagini \cite{URN}, e un sistema generico di {\sl label} associate agli {\sl edge} del grafo primario.
In questo modo il problema dell'identificazione della {\sl space syntax} viene spostato sui metodi di identificazione delle {\sl label} e quindi sulla misura di generiche variabili urbanistiche, non necessariamente legate alle linee assiali.
L'algoritmo proposto nei prossimi paragrafi ritrova la {\sl space syntax} ``tradizionale'' come caso particolare, sanando anche alcune delle apparenti incongruenze, ma amplia anche la teoria includendo fenomeni urbani che non sono direttamente connessi a esperienze visuali.

Come applicazione del metodo viene costruita una {\sl space syntax} acustica, cio\`e una mappa topologica della citt\`a che rappresenta lo spazio urbano sulla base delle sue qualit\`a acustiche, propriet\`a che possono essere percepite da un non vedente, o anche solo dall'udito di chi si muove nella citt\`a. 

Il lavoro \`e organizzato nel modo seguente.
Nel prossimo paragrafo viene data una breve descrizione dei risultati e dei metodi della {\sl space syntax}.
Nel paragrafo successivo verranno introdotte le definizioni di grafo primario $G_p$ etichettato e del corrispondente {\sl line graph} contratto.
Viene poi studiato un caso elementare per esemplificare il funzionamento del metodo.
Infine viene analizzato un caso reale in cui le etichette degli {\sl edge} rappresentano il livello di inquinamento acustico di un quartiere di Roma.
Nelle conclusioni sono indicate alcune possibili vie di sviluppo di questi metodi.

\subsection{La space syntax}
La definizione che Hillier, Hanson e Graham danno nel 1987 della {\sl space syntax} \cite{RePEc:pio:envirb:v:14:y:1987:i:4:p:363-385}, riportata in \cite{Ratti2004}, evidenzia come alla base della teoria ci siano le configurazioni spaziali della citt\`a, ma non indica alcuna grandezza che sia rappresentativa di queste configurazioni:\\\\
{\sl ``Space syntax ... is a set of techniques for the representation, quantification, and interpretation of spatial configuration in buildings and settlements.
Configuration is defined in general as, at least, the relation between two spaces taking into account a third, and, at most, as the relations among spaces in a complex taking into account all other spaces in the complex.
Spatial configuration is thus a more complex idea than spatial relation, which need invoke no more than a pair of related spaces.''}\\\\
Nello sviluppo della {\sl space syntax} l'esperienza visuale, come \`e naturale, \`e stata dominante.
Il modello privilegia la percezione visiva degli spazi, ma gli spazi urbani possono essere percepiti anche con l'udito, con altri sensi e in generale le loro caratteristiche possono essere misurate con uno strumento.
Questo approccio da una parte \`e stato molto proficuo per la comprensione di come viene percepita la citt\`a da chi la attraversa, dall'altra ha limitato in qualche senso il quadro urbanistico.

Una persona che si muove nella citt\`a la ``sente'' anche come suono, come odore, come densit\`a di folla, come pericolo, come riferimenti culturali o come il risultato del comportamento di altre variabili non immediatamente identificabili con la vista.
L'aspetto visuale \`e certamente dominante, ma le prospettive spaziali sono solo una delle possibili variabili che identificano gli spazi.
L'analisi di \cite{Ratti2004} \`e forse un po' condizionata da questo punto di vista nella critica alla poca robustezza del metodo delle linee assiali rispetto a piccole modifiche della struttura urbana.
In \cite{Ratti2004} viene infatti sottolineato che le propriet\`a topologiche della {\sl space syntax} non colgono alcuni aspetti della citt\`a che sono connessi alle propriet\`a metriche e viene messo in evidenza che le linee assiali possono spezzarsi se si deforma leggermente il tessuto urbano.

\`E quindi necessaria una corretta e pi\`u generale definizione delle linee assiali, o dei nodi dei corrispondenti grafi, che porti ad una generalizzazione della {\sl space syntax} e alla possibilit\`a di considerare anche variabili e grandezze che non sono di tipo visuale.

La perdita di riferimenti spaziali \`e conseguenza del fatto che le linee assiali non possono essere collocate in un punto dello spazio.
Pi\`u in generale, se geolocalizziamo un grafo per mezzo delle coordinate dei suoi nodi, il corrispondente {\sl line graph} non sar\`a comunque geolocalizzato.
I {\sl line graph} danno informazioni sulla connettivit\`a dello spazio, ma perdono le possibili propriet\`a metriche del grafo primario.

Tuttavia, in contrasto con quanto suggerito in \cite{Ratti2004}, non tutti gli aspetti della rappresentazione topologica del reticolo urbano sono stati approfonditi e il modello, come vedremo, se opportunamente formalizzato, può essere ancora fonte di utili informazioni.
Inoltre la connettivit\`a non deve essere necessariamente spaziale, la citt\`a pu\`o connettere servizi, comportamenti, emergenze e ciascuna variabile determina una propria connettivit\`a.

Resta il problema di come definire il sistema di etichette, quindi la scelta delle variabili definite sugli spazi aperti, ma questo problema, di pertinenza degli urbanisti, \`e un problema di scelta delle variabili da studiare e delle metodologie per misurarle.
Le linee assiali sono facilmete identificabili dalla geometria dello spazio, ma possono anche essere misurate direttamente negli spazi reali.
Partendo da questa considerazione ogni propriet\`a direttamente misurabile di uno spazio aperto \`e una variabile definita su un grafo primario.
\`E possibile quindi misurare il livello della criminalit\`a, dei servizi, la densit\`a abitativa, i prezzi immobiliari e osservare che lo spazio urbano mette in relazione i diversi valori di queste variabili.

\section{I grafi urbani}
Proviamo allora a ridefinire la teoria seguendo e generalizzando il lavoro di Crucitti, Latora e Porta \cite{crucitti2006centrality}, cercando anche di riordinare la terminologia sull'argomento che in letteratura \`e un po' disordinata.
In \cite{BV_Buch} per esempio viene usato il termine ``duale'' per identificare il grafo assiale della {\sl space syntax}, ma nel linguaggio della teoria dei grafi la dualità indica un altro tipo di trasformazione di un grafo \cite{Diestel2005}.
Inoltre a differenza dei grafi spazialmente duali, i grafi assiali non sono involutivi.\\

\begin{figure}[!ht]
\centering
\includegraphics[width=0.6\textwidth]{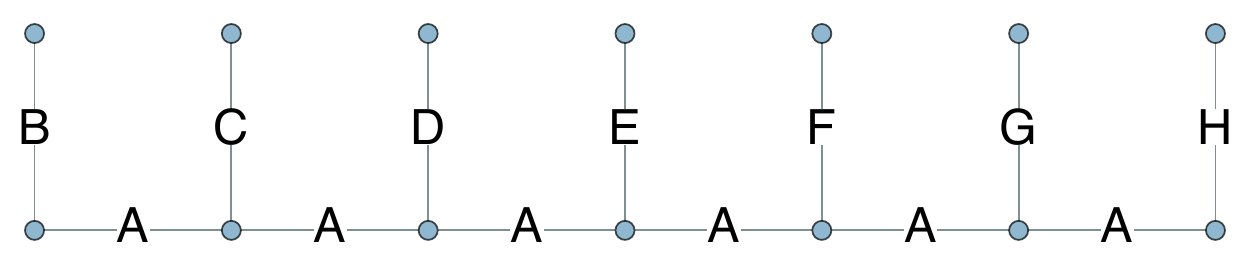}
\caption{Il grafo di un fenomeno di {\sl urban sprawl}, la strada centrale $A$ \`e collegata ai vicoli ciechi $B,C,D,E,F,G,H$}
\label{fig4}
\end{figure}

Definiamo {\bf grafo primario geolocalizzato} o semplicemente {\bf grafo primario} $G_p$, un grafo i cui {\sl edge} sono gli spazi aperti di una mappa urbana e i nodi sono gli incroci tra questi spazi.
Ad ogni nodo sono quindi associate le sue coordinate spaziali.
Per esempio possiamo pensare che una strada che ne incrocia altre (fig.\ref{fig4}) sia rappresentata da un {\sl edge}, mentre i due incroci ai suoi estremi saranno nodi.

Questo tipo di rappresentazione, per quanto intuitiva e immediata, non \`e per\`o univoca.
Per esempio una piazza circolare con una fontana al centro pu\`o essere rappresentata come un ciclo chiuso o come un unico nodo.
Un opportuno sistema di etichette sugli {\sl edge} di $G_p$, insieme al processo di contrazione che verr\`a definito, pu\`o per\`o rendere coerenti diversi grafi primari che rappresentano gli stessi spazi aperti.
Il grafo primario appena descritto pu\`o essere estratto dalla mappa di una citt\`a anche per mezzo di un semplice procedimento di analisi dell'immagine \cite{URN}.
Diamo quindi la seguente definizione.\\

\begin{definition}[Grafo Primario Geolocalizzato]\ \\
	Un grafo primario geolocalizzato $G_p$ \`e la coppia $(V,E)$ dove $V$ \`e un insieme di coordinate geografiche in $\mathbb{R}^2$, mentre $E\subseteq V\times V$.
	Inoltre su $E$ \`e definita una mappa $c:E\rightarrow C$ che associa un'etichetta di un generico insieme $c\in C$ ad ogni edge $e\in E$.
	\label{gue}
\end{definition}
\ \\
Nel seguito tratteremo solo grafi non diretti.
In genere $G_p$ \`e quasi planare, nel senso che possono esserci pochi {\sl edge} che corrispondono a ponti o a strutture sovrapposte che ne alterano la planarit\`a.
Inoltre due nodi con le stesse coordinate possono essere distinti se utilizziamo le coordinate in $\mathbb{R}^3$ invece delle coordinate in $\mathbb{R}^2$.
In questo caso due nodi che hanno le stesse coordinate spaziali $(x,y,z)$ sono lo stesso nodo.
Per semplicit\`a di rappresentazione assumiamo che le coordinate di $G_p$ siano in $\mathbb{R}^2$. 

Ad ogni {\sl edge} di $G_p$ associamo una variabile a valori in $C$, che misura una propriet\`a urbanistica.
Naturalmente questa non \`e una colorazione del grafo perch\'e {\sl edge} adiacenti possono avere lo stesso colore.
Nel caso particolare della {\sl space syntax} le etichette, o i colori, rappresentano le linee assiali, ma, per come abbiamo definito $G_p$, la stessa linea assiale pu\`o esser spezzettata in {\sl edge} che hanno la stessa etichetta comr mostrato in fig.\ref{fig4}.

Come abbiamo detto, le etichette possono indicare per esempio i nomi delle strade, il tipo di strada, la portata, le linee assiali, il livello di inquinamento o qualsiasi altra caratteristica urbana misurabile ed associabile a un {\sl edge}.
Per costruire il {\bf grafo urbano} corrispondente al grafo assiale nel caso particolare in cui le etichette identificano una stessa linea di vista, abbiamo bisogno della nozione di {\sl line graph} \cite{Diestel2005}.\\

\begin{figure}[!ht]
\centering
\includegraphics[width=0.6\textwidth]{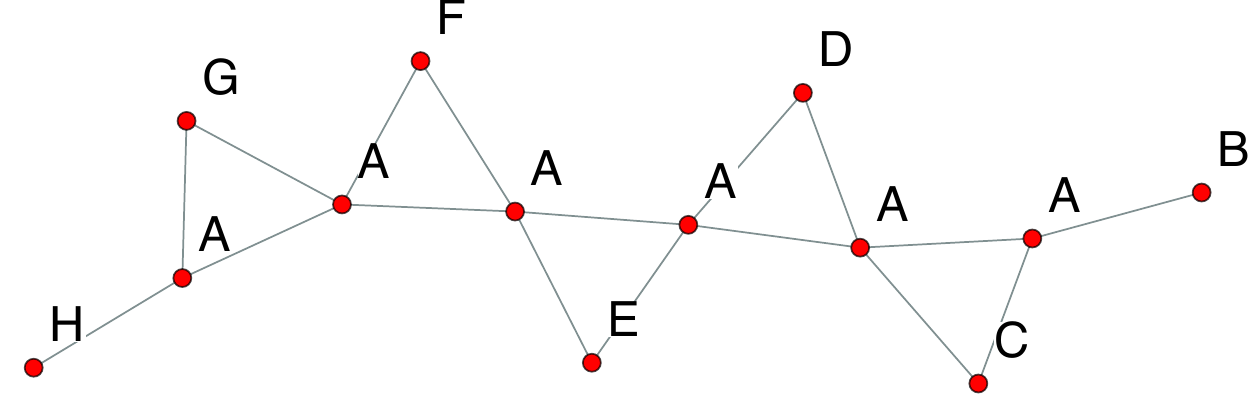}
\caption{Il {\sl line graph} del grafo di fig.\ref{fig4}}
\label{fig5}
\end{figure}

\begin{definition}[Line Graph]
	Dato un grafo $G=(V,E)$ il corrispondente {\sl line graph} $L(G)$ \`e il grafo su $E$ tale che due nodi di $L(G)$ sono connessi se i due corrispondenti {\sl edge} sono adiacenti in $G$.
\end{definition}
\ \\
\`E evidente che se anche $G$ \`e un grafo planare, il corrispondente {\sl line graph} non sar\`a in generale un grafo planare.
Nel {\sl line graph} si perdono quindi tutte le caratteristiche spaziali del grafo di partenza, ma si ottengono informazioni sulla topologia e sulla connettivit\`a dello spazio urbano che viene rappresentato.
Ad esempio un nodo di $G$ di grado $d$ diventa una {\sl clique} di $L(G)$ cio\`e un sottoinsieme di un grafo in cui tutti i nodi sono reciprocamente connessi.
Il {\sl line graph} di $G_p$ risulta quindi pi\`u interessante, urbanisticamente, del grafo primario, perch\'e possiede una maggiore variet\`a di distribuzione del grado.\\\\
Quando costruiamo il {\sl line graph} $L(G_p)$ di un grafo urbano con etichette, le etichette degli {\sl edge} di $G_p$ vengono associate ai corrispondenti nodi di $L(G_p)$.
Otteniamo quindi un nuovo grafo, non planare e con i nodi etichettati.
Questo grafo non \`e, a parte casi particolari, lo stesso grafo assiale della {\sl space syntax} neanche quando le etichette indicano correttamente le linee assiali.
Per esempio nel caso di un fenomeno di {\sl urban sprawl} (fig.\ref{fig4}) non si ottiene una stella come si otterrebbe utilizzando \cite{VaroudisX}, ma un grafo in cui la continuit\`a della strada principale $A$ \`e spezzettata in tanti vertici adiacenti con la stessa etichetta (fig.\ref{fig5}).
\ \\\\

\begin{figure}[!ht]
\centering
\includegraphics[width=0.6\textwidth]{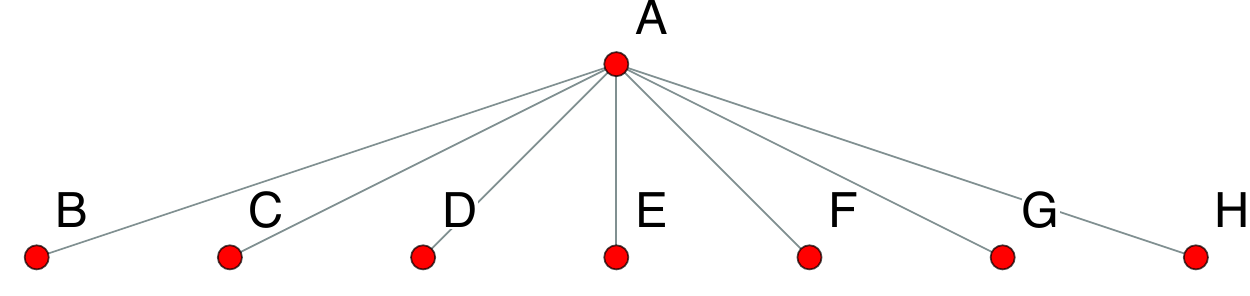}
\caption{Il grafo a stella che si ottiene contraendo {\sl line graph} del grafo di fig.\ref{fig5}}
\label{fig6}
\end{figure}

\ \\\\
\ \\\\\\
Possiamo allora introdurre un opportuno procedimento di contrazione sul {\sl line graph} e utilizzare il grafo contratto come definizione di {\bf grafo urbano}.
La contrazione \cite{Diestel2005} di un grafo su un {\sl edge} \`e il grafo nel quale quell'{\sl edge} (e quindi quella coppia di nodi) \`e stato sostituito da un nuovo vertice che eredita tutti gli {\sl edge} dei due nodi eliminati.
Formalmente la contrazione $G\setminus(x,y)$ di un grafo $G$ sull'{\sl edge} $(x,y)$ \`e definita nel modo seguente.\\
\begin{definition}[Contrazione]
	Dato un grafo $G=(V,E)$ e dati due vertici $x$ e $y$, il grafo contratto sull'{\sl edge} $(x,y)$, $G_c=G\setminus(x,y)=(V_c,E_c)$ \`e il grafo con vertici $V_c=\big(E\setminus\{x,y\}\big)\cup v$ dove $v$ \`e un nuovo vertice, $v\notin V\cup E$, e $E_c= E\setminus(x,y)\cup\{(v,z)| (x,z)\in E \text{ o } (y,z)\in E\}$. 
\end{definition}\ \\
Le informazioni che un {\sl line graph} contratto d\`a sulla connettivit\`a degli spazi urbani, sono estremamente importanti dal punto di vista urbanistico e perfettamente coerenti, quando le etichette corrispondono alle linee assiali, con i risultati della {\sl space syntax}.
Ad esempio una strada con tante uscite secondarie viene trasformata dalla contrazione di fig.\ref{fig2} o dai metodi dei grafi assiali, in una topologia a stella \cite{BV_Buch} che identifica un fenomeno di {\sl urban sprawl} (fig.\ref{fig6}) in cui la chiusura della strada $A$ elimina ogni possibile connessione tra le altre strade.
Definiamo allora il {\bf grafo urbano}, che generalizza il concetto di grafo assiale.\\

\begin{definition}[Grafo Urbano]
	Dato un {\sl line graph} $L(G_p)$ di un grafo primario geolocalizzato, il grafo urbano si ottiene contraendo i vertici adiacenti che hanno la stessa etichetta.
	Il nuovo nodo $v$ ottenuto dalla contrazione di $(x,y)$ avr\`a la stessa etichetta dei due nodi. 
	Gli {\sl edge} multipli vengono trasformati in un unico {\sl edge}.
\end{definition}
\ \\
In altri termini possiamo raggruppare gli spazi aperti in classi determinate dallo stesso valore di una variabile urbanistica e dall'adiacenza dei nodi del corrispondente {\sl line graph}.
Le operazioni di contrazione e di {\sl line graph} non commutano: se contraiamo su $G_p$ gli {\sl edge} adiacenti con la stessa etichetta e poi costruiamo il {\sl line graph}, otteniamo un grafo diverso dal grafo contratto di $L(G_p)$.
Nel caso dell'{\sl urban sprawl} di fig.\ref{fig1} la non commutativit\`a \`e evidente.\\\\
Osserviamo anche che le interazioni tra i nodi del grafo contratto continuano ad essere determinate da intersezioni, ma non pi\`u dalle intersezioni delle linee assiali, ma da quelle di generici spazi determinati dalla classe di equivalenza dell'etichetta, e quindi dalle classi di equivalenza di generiche grandezze urbanistiche.
\ \\\\
\ \\\\
Utilizzando la definizione di grafo urbano possiamo scrivere alcuni moduli con Mathematica per costruire $L(G_p)$ e per studiarne le caratteristiche di interesse urbanistico.\\\\
\section{La costruzione del grafo urbano con Mathematica}
Per realizzare il {\bf grafo urbano} con Matematica \`e necessario implementare alcune funzioni che riportiamo in modo schematico, sottolineando che per chiarezza non sono ottimizzate dal punto di vista della velocit\`a di calcolo.

Per prima cosa abbiamo bisogno di una funzione {\tt Contrazione[grafo, n1, n2, nomi]} che dato un {\tt grafo}, e un insieme di etichette {\tt nomi}, contrae i due nodi {\tt n1} e {\tt n2} e restituisce la nuova matrice di adiacenza e la nuova lista di etichette.\\
	\begin{lstlisting}[frame=single]  % Start your code-block
		
Contrazione[grafo_,n1_,n2_,nomi_]:=Module[
	{ma,ma2,ma3,ma4,nomi2},
	ma=Normal[AdjacencyMatrix[grafo]];
	ma2=ma;
	ma2[[All,n1]]+=ma2[[All,n2]];
	ma2[[n1,All]]+=ma2[[n2,All]];
	ma3=Drop[Clip[ma2],{n2,n2},{n2,n2}];
	ma4=ma3-DiagonalMatrix[Diagonal[ma3]];
	nomi2=Drop[nomi,{n2,n2}];
	Return[{ma4,nomi2}]
]	
	\end{lstlisting}
	\ \\La funzione {\tt Contrazione[grafo,n1,n2,nomi]} d\`a il grafo sotto forma di una matrice di adiacenza, ma può essere pi\`u comodo avere una funzione {\tt Contrazione2[grafo,n1,n2,nomi]} che dia direttamente il grafo e il sistema di etichette:\\
	\begin{lstlisting}[frame=single]
Contrazione2[grafo_,n1_,n2_,nomi_]:=Module[
	{ma,ma2,ma3,ma4,nomi2,gc,grafocontratto},
	ma=Normal[AdjacencyMatrix[grafo]];
	ma2=ma;
	ma2[[All,n1]]+=ma2[[All,n2]];
	ma2[[n1,All]]+=ma2[[n2,All]];
	ma3=Drop[Clip[ma2],{n2,n2},{n2,n2}];
	ma4=ma3-DiagonalMatrix[Diagonal[ma3]];
	nomi2=Drop[nomi,{n2,n2}];
	gc=AdjacencyGraph[ma4];
	grafocontratto=SetProperty[gc,
		VertexLabels->Table[i->nomi2[[i]],
			{i,1,Length[nomi2]}]];
	Return[{grafocontratto,nomi2}]
]
	\end{lstlisting}

	\ \\Abbiamo poi bisogno di una funzione {\tt NodiAdiacenti[ilgrafo,inomi]} che legge un grafo e una lista di etichette sui nodi, individua i nodi con lo stesso nome e restituisce i primi due nodi adiacenti con la stessa etichetta:\\
	\begin{lstlisting}[frame=single]
NodiAdiacenti[ilgrafo_,inomi_]:=Module[
{coppie,indice,posizione,n1,n2},
coppie=Table[{inomi[[EdgeList[ilgrafo][[i,1]]]],
	inomi[[EdgeList[ilgrafo][[i,2]]]]},{i,1,Length[EdgeList[ilgrafo]]}];
indice=Table[Length@Tally@coppie[[i]],{i,1,Length[coppie]}];
If[MemberQ[indice,1],posizione=Position[indice,1][[1,1]],posizione=0];
Return[{EdgeList[ilgrafo][[posizione,1]],EdgeList[ilgrafo][[posizione,2]]}]
]
	\end{lstlisting}
\ \\ Per operare la contrazione introduciamo la funzione {\tt ContrazioneGrafo[ilgrafo,inomi]} che individua due nodi adiacenti con la stessa etichetta e li contrae:\\
\begin{lstlisting}[frame=single]
ContrazioneGrafo[ilgrafo_,inomi_]:=Module[
	{a,ilgrafod,inomid,n1,n2},
	ilgrafod=ilgrafo;
	inomid=inomi;
	If[Length[NodiAdiacenti[ilgrafo,inomi]]>1,
		n1=NodiAdiacenti[ilgrafo,inomi][[1]];
		n2=NodiAdiacenti[ilgrafo,inomi][[2]];
		{ilgrafod,inomid}=Contrazione2[ilgrafo,n1,n2,inomi];
	];
	Return[{ilgrafod,inomid}]
]
\end{lstlisting}
\ \\Il modulo {\tt ContraibileQ[ilgrafo,inomi]}, verifica la contraibilit\`a del grafo:\\
\begin{lstlisting}[frame=single]
ContraibileQ[ilgrafo_,inomi_]:=Module[
	{coppie,indice,posizione},
	coppie=Table[
		{inomi[[EdgeList[ilgrafo][[i,1]]]],
		inomi[[EdgeList[ilgrafo][[i,2]]]]},
		{i,1,Length[EdgeList[ilgrafo]]}
		];
		indice=Table[Length@Tally@coppie[[i]],{i,1,Length[coppie]}];
		Return[MemberQ[indice,1]]
]
\end{lstlisting}
\ \\La funzione {\tt ilGrafoPrimario[regole]} costruisce $G_p$ da un insieme di regole, cio\`e dalle coppie ({\sl edge}, etichetta)  sugli {\sl edge}. Per semplicit\`a non associamo anche le coordinate spaziali dei nodi:\\
\begin{lstlisting}[frame=single]
GrafoEtichette[regole_] := Module[
	{g, e},
	g = regole /. Rule -> (#1 &);
	e = regole /. Rule -> (#2 &);
	Return[{Graph[g, EdgeLabels -> regole, 
		EdgeLabelStyle -> Directive[Red], VertexLabels -> "Name"], e}]
]
\end{lstlisting}
\ \\La funzione {\tt LineGraphID[regole]} costruisce il {\sl line graph} da un insieme di regole, etichettandone i nodi con le etichette degli {\sl edge} di $G_p$:\\
\begin{lstlisting}[frame=single]
LineGraphID[regole_] := Module[
	{a, b, e},
	a = ilGrafoPrimario[regole];
	b = LineGraph[a];
	e = regole /. Rule -> (#2 &);
	Return[SetProperty[b, 
		VertexLabels -> Table[i -> e[[i]], {i, 1, Length[e]}]]]
]
\end{lstlisting}
\ \\Infine la funzione {\tt LineGraphIdContratto[regole]} legge un insieme di regole e restituisce il LineGraph contratto:\\
\begin{lstlisting}[frame=single]
LineGraphIDContratto[regole_] := Module[
	{ilgrafo, inomi, risultato},
	ilgrafo = LineGraphID[regole];
	inomi = regole /. Rule -> (#2 &);
	Return[ContrazioneCompletaGrafo[ilgrafo, inomi]]
]
\end{lstlisting}
\ \\
Questi sono, in modo schematico, i moduli principali che possono essere utilizzati per realizzare i grafi urbani.
Nel paragrafo successivo questi metodi verranno utilizzati per studiare un caso urbanistico reale.

\section{La space syntax acustica}
\begin{figure}[!ht]
\centering
\includegraphics[width=0.6\textwidth]{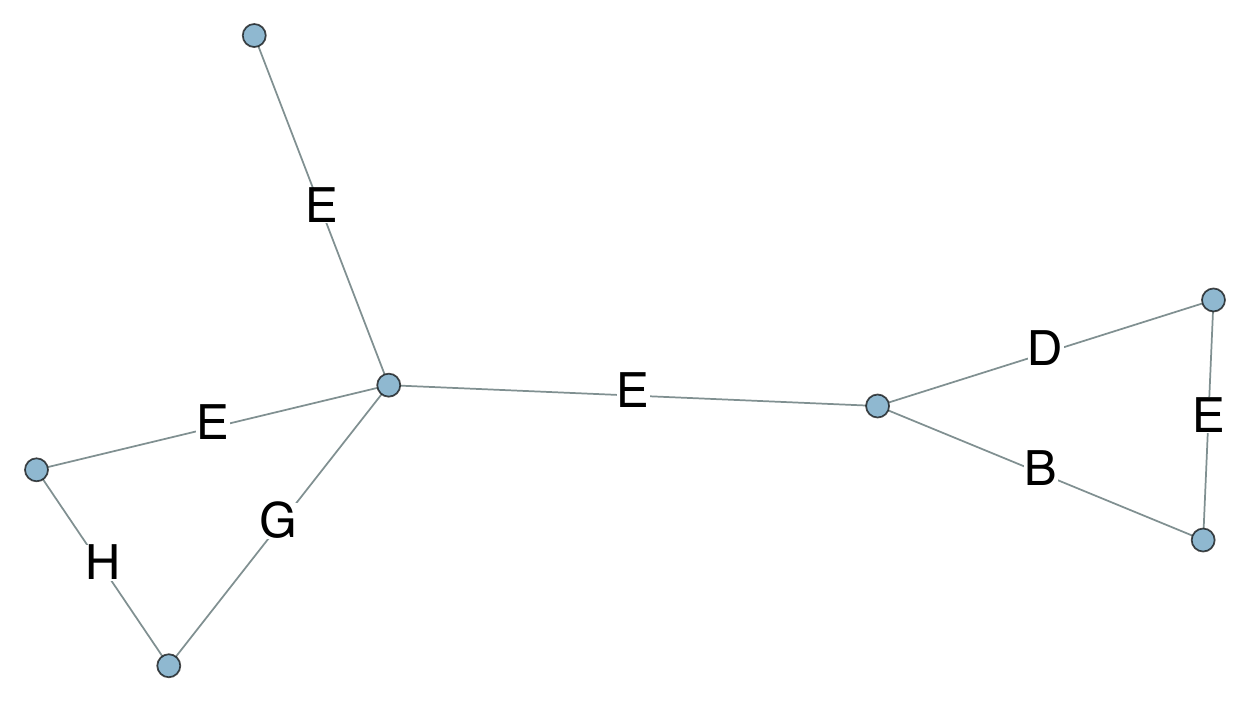}
\caption{Il grafo primario con i nodi geolocalizzati e un sistema di etichette sugli {\sl edge}}
\label{fig1}
\end{figure}

\begin{figure}[!ht]
\centering
\includegraphics[width=0.6\textwidth]{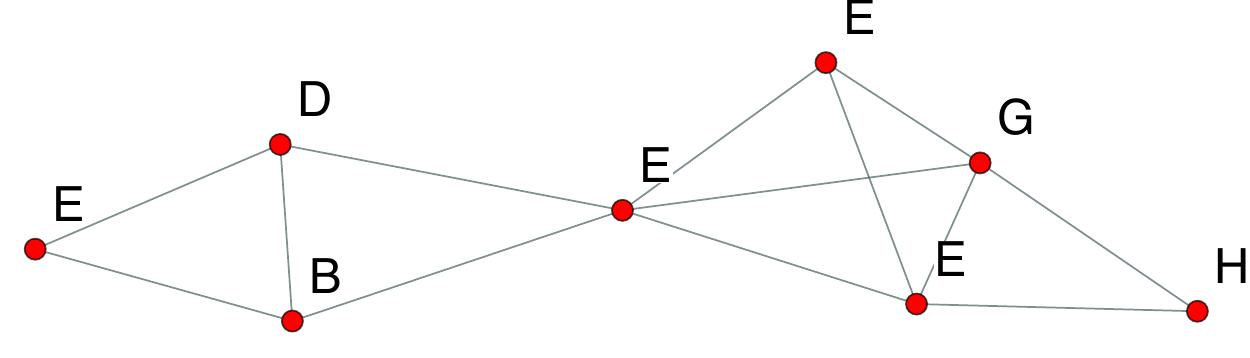}
\caption{Il {\sl line graph} di fig.\ref{fig1} (non contratto)}
\label{fig2}
\end{figure}

\begin{figure}[!ht]
\centering
\includegraphics[width=0.6\textwidth]{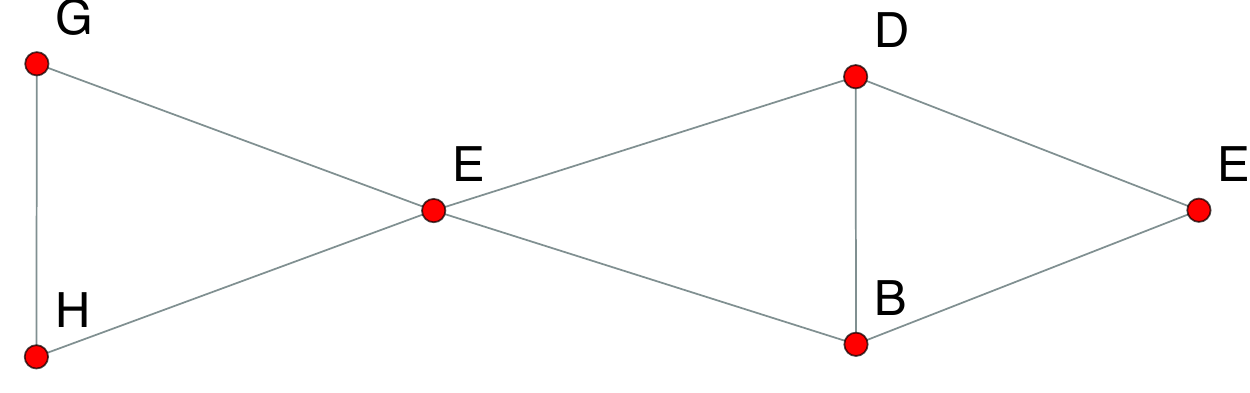}
\caption{Il {\sl line graph} di fig.\ref{fig1} dopo la contrazione}
\label{fig3}
\end{figure}

Per comprendere come funziona il metodo appena definito, possiamo costruire il grafico di fig.\ref{fig1} dal quale, per mezzo della funzione {\tt LinegraphID} otteniamo il corrispondente {\sl line graph} con le etichette sui vertici di fig.\ref{fig2}.
La funzione {\tt LinegraphIDContratto} restituisce il grafo finale fig.\ref{fig3}, dove possono esserci nodi con la stessa etichetta, ma non adiacenti.

Come nel caso dell'{\sl urban sprawl} il risultato \`e lo stesso che otterremmo con i grafi assiali se identificassimo con un'unica linea di vista gli {\sl edge} adiacenti con la stessa etichetta.
Il sistema di etichette pu\`o essere un insieme qualsiasi, ma deve avere una coerenza nel senso che l'etichetta deve rappresentare una propriet\`a dell'{\sl edge}.
Esempi di etichette coerenti sono sono le linee assiali, l'inquinamento acustico, la densit\`a abitativa, ma non le misure relative agli {\sl edge} adiacenti, come per esempio il senso di marcia di una strada.
Inoltre la relazione tra due nodi determinata dalla propriet\`a di possedere la stessa etichetta \`e una relazione di equivalenza.

\begin{figure}[!h]
\centering
\includegraphics[width=0.51\textwidth]{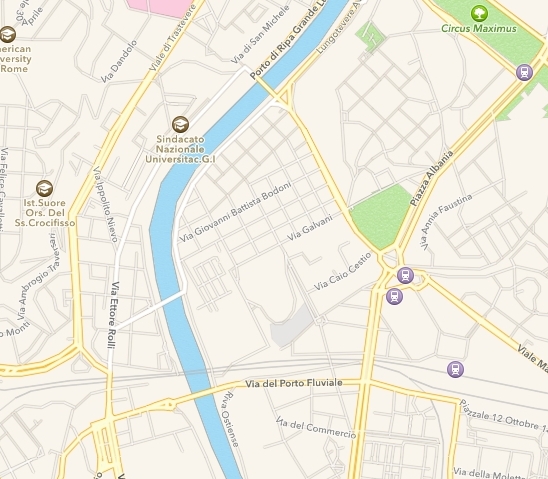}
\caption{L'area di studio, il quartiere Testaccio di Roma}
\label{testaccio}
\end{figure}

Possiamo allora esaminare come caso reale il grafo che si ottiene da un'analisi del livello di inquinamento acustico del quartiere Testaccio di Roma di fig.\ref{testaccio}.
Le misure dell'intensit\`a del rumore urbano in {\sl decibel} sono state fatte alla stessa ora in giorni diversi e non sono molto indicative della situazione acustica del quartiere, ma sono utili per mostrare le possibilit\`a del metodo introdotto.
In un lavoro successivo introdurremo una metodologia di misura acustica molto precisa basata sulla dinamica dei suoni, e studiamo l'effettivo funzionamento dell'acustica dell'area con il metodo proposto.

\begin{figure}[!h]
\centering
\includegraphics[width=0.5\textwidth]{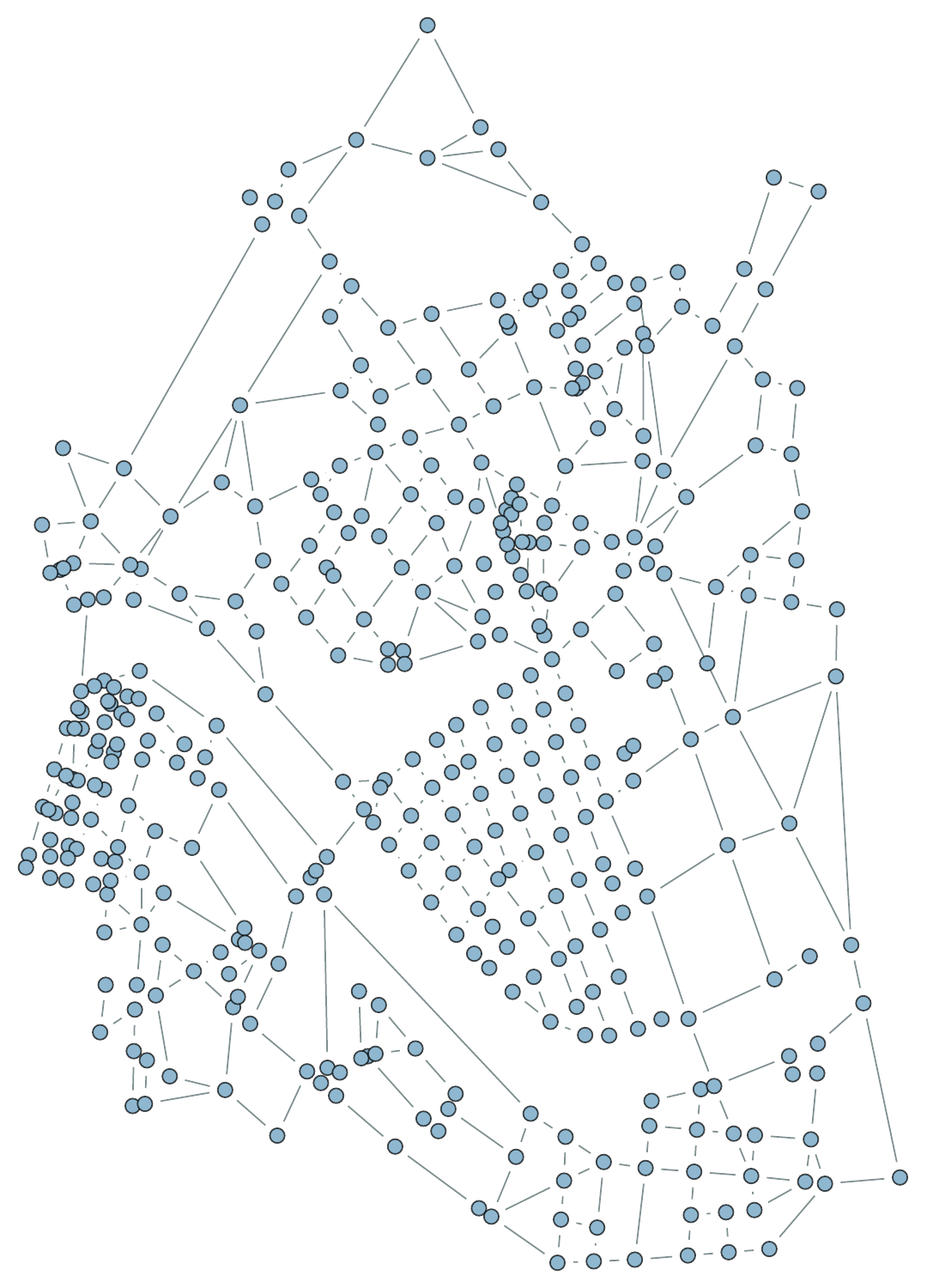}
\caption{Il grafo dell'area di studio}
\label{grafotestaccio}
\end{figure}

\begin{figure}[!h]
\centering
\includegraphics[width=0.5\textwidth]{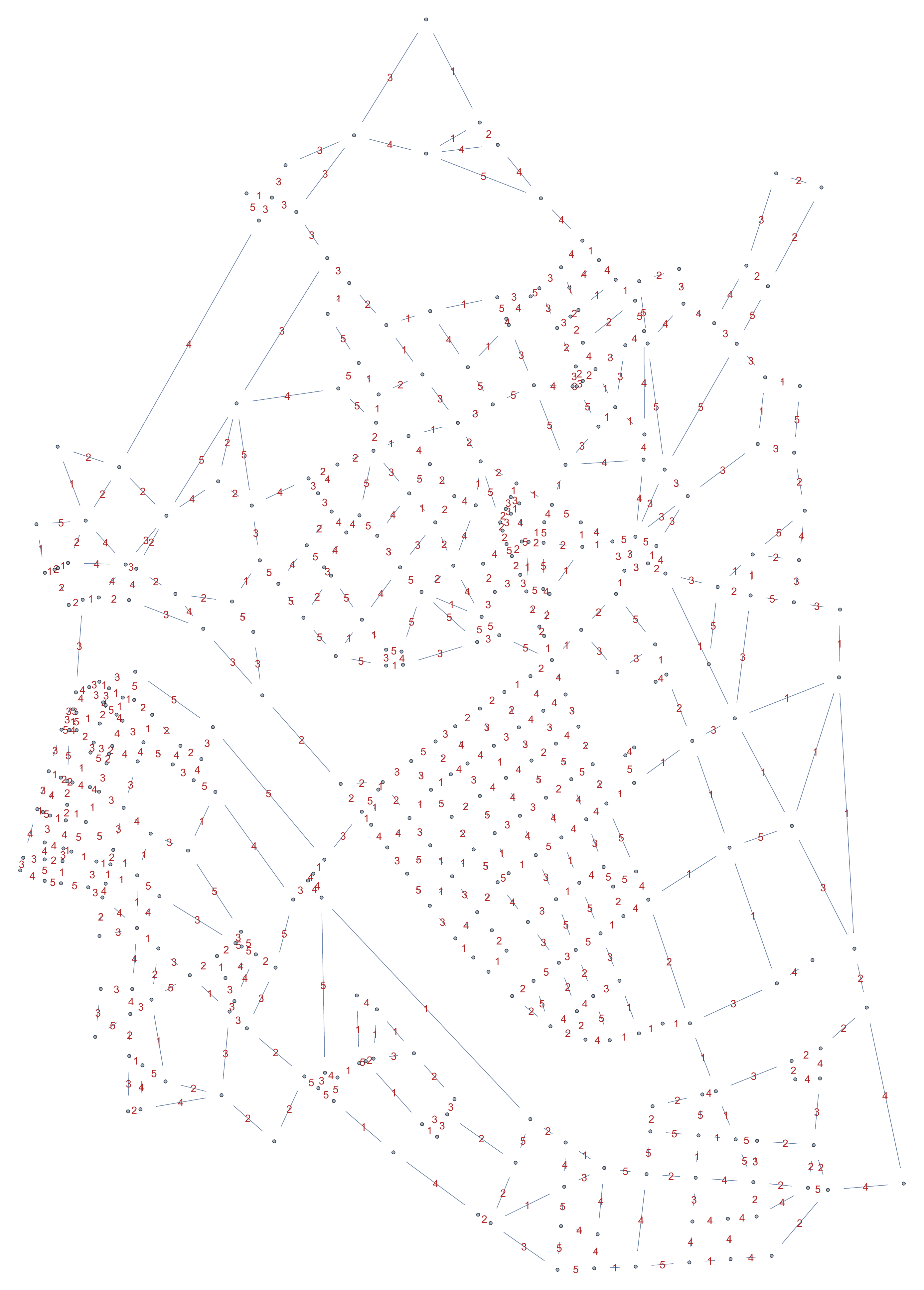}
\caption{Il grafo dell'area di studio con le misure di inquinamento acustico}
\label{grafotestacciolabel}
\end{figure}

I dati utilizzati nell'esempio non sono mediati su un'intervallo di tempo, ma hanno solo un valore illustrativo.
Il grafo primario $G_p$ che si ricava dalla mappa di fig.\ref{testaccio} \`e rappresentato in fig.\ref{grafotestaccio} mentre lo stesso grafo etichettato con le misure acustiche \`e riportato in fig.\ref{grafotestacciolabel}.
In entrambi i grafi i nodi sono geolocalizzati.

\begin{figure}[!ht]
\centering
\includegraphics[width=0.6\textwidth]{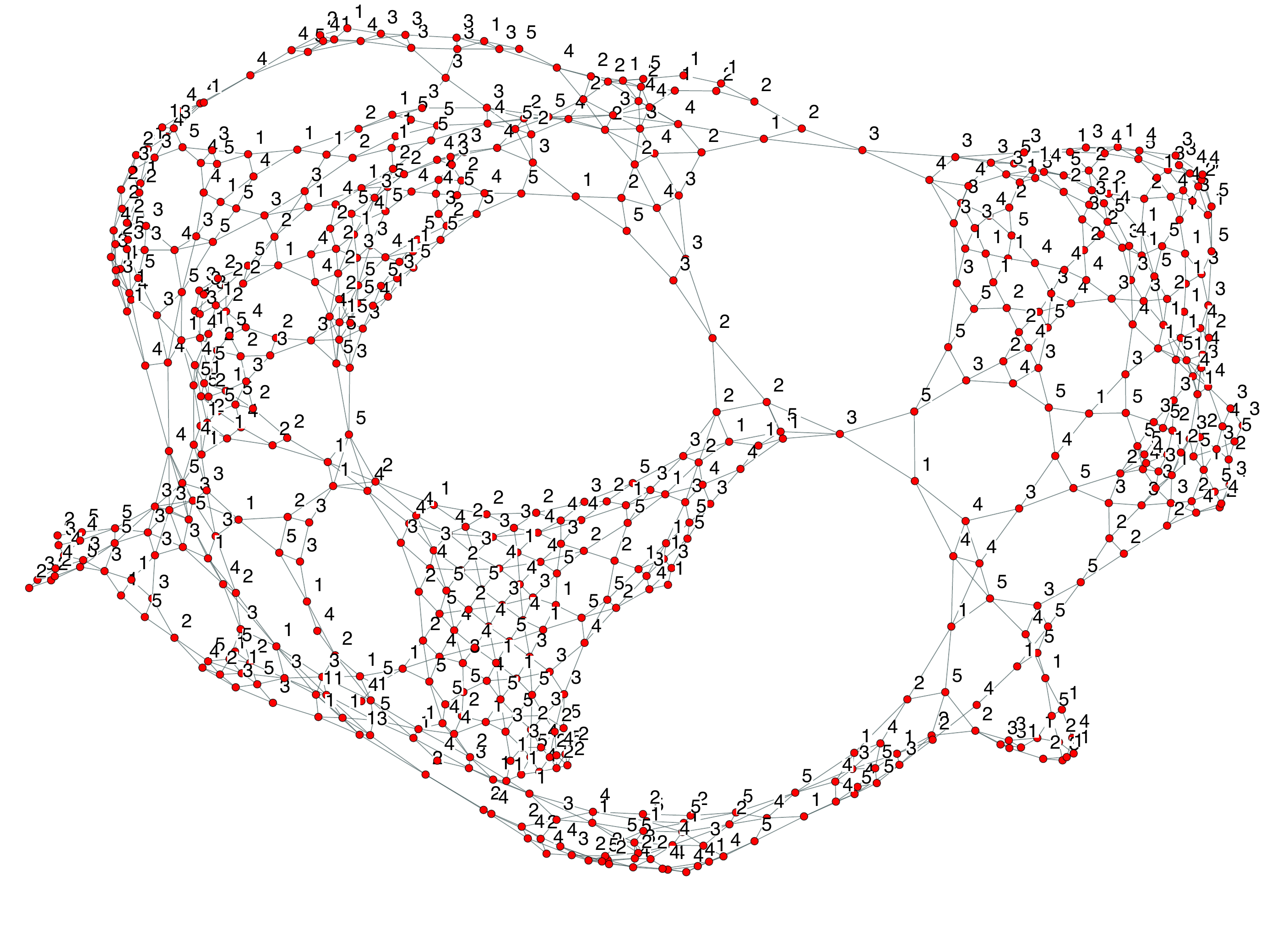}
\caption{Il {\sl line graph} dell'area di studio con le etichette sui vertici}
\label{testaccioLineGraph}
\end{figure}

\ \\
Una misura del livello di inquinamento acustico associa la variabile $c\in C\equiv\{1,2,3,4,5\}$ ad ogni {\sl edge}, e indica un livello di inquinamento acustico misurato in decibel e discretizzato su una scala di cinque valori (fig.\ref{grafotestacciolabel}).
Se costruiamo il {\sl line graph} non contratto del grafo primario con le etichette, otteniamo il grafo di fig.\ref{testaccioLineGraph}.
I nodi ereditano le misure acustiche degli {\sl edge} corrispondenti e il grafo perde ogni riferimento alle coordinate spaziali.
La rappresentazione del livello di inquinamento acustico dal punto di vista della {\sl space syntax} d\`a importanti informazioni sull'accessibilit\`a ``acustica'' degli spazi urbani e introduce un metodo per descrivere la zonizzazione acustica di un'area urbana.
Le grandezze caratteristiche della {\sl space syntax} \cite{Volchenkov2008,GlossarySS}, come per esempio la profondit\`a o l'asimmetria relativa, possono essere facilmente generalizzate al caso dei grafi acustici.

Per fare un esempio, dati due nodi $i$ e $j$ del grafo urbano, quindi due aree urbane connesse e con lo stesso livello di inquinamento acustico, la loro profondit\`a relativa $d_{ij}$ ({\sl depth}) \`e  il pi\`u piccolo numero di passi necessari per andare da $i$ a $j$ \cite{GlossarySS}.
Sommando le profondit\`a di un nodo $i$ rispetto a tutti gli altri otteniamo la profondit\`a totale ({\sl total depth}) del un nodo $i$:
\begin{equation}
	{\cal D}_i=\sum_{j=1}^Nd_{ij}
\end{equation}
dove $N$ \`e il numero di nodi del grafo.
Il suo valore medio ({\sl mean depth})
\begin{equation}
	l_i={{\cal D}_i\over N-1}
\end{equation}
viene chiamato profondit\`a media ed \`e una misura del grado di integrazione/segregazione del nodo $i$-esimo.
Nel caso acustico, $l_i$ misura il livello di integrazione nel tessuto urbano di un'area di maggiore o minore rumore.
\begin{figure}[!ht]
\centering
\includegraphics[width=0.6\textwidth]{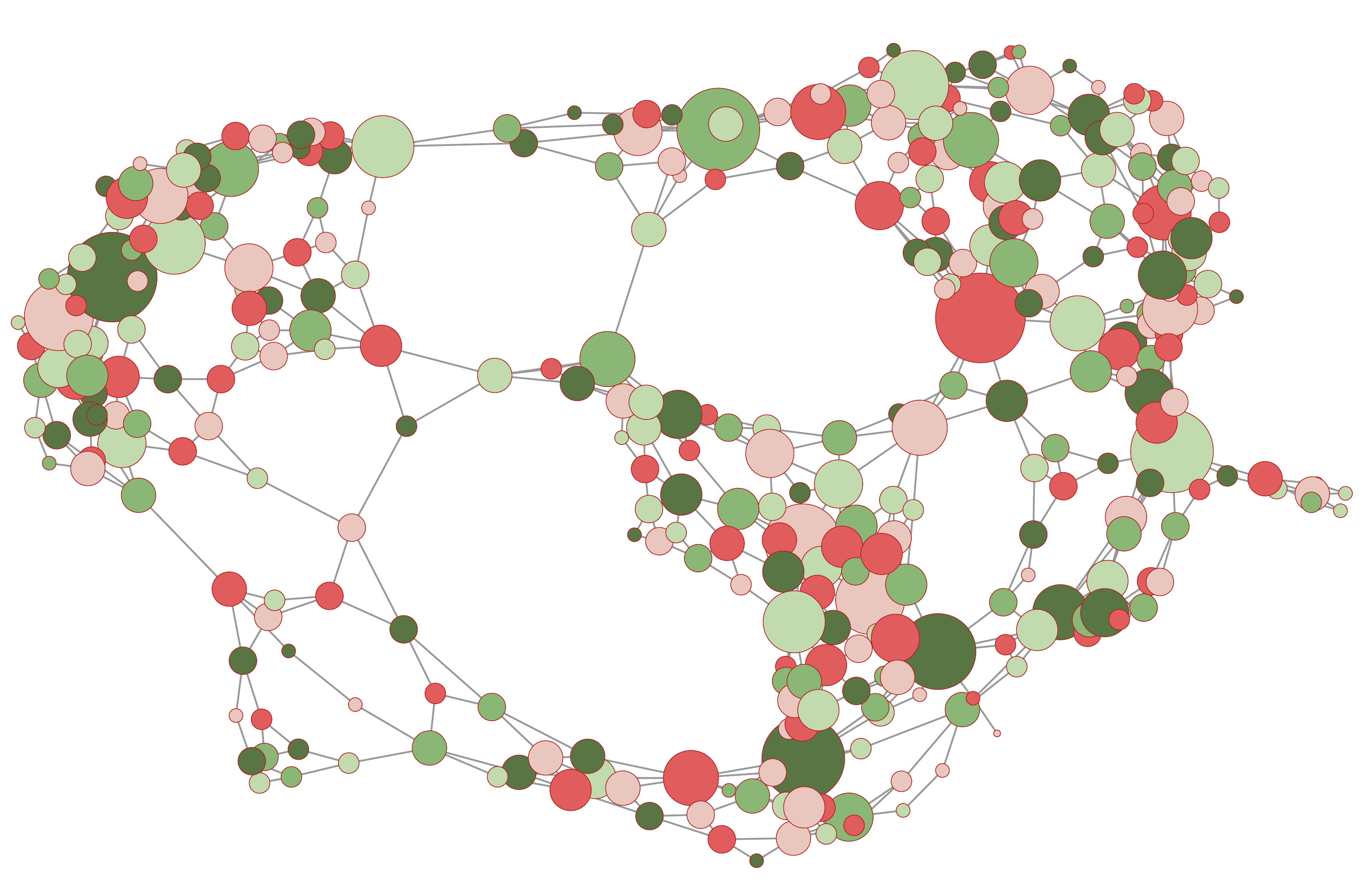}
\caption{Il grafo urbano acustico dell'area di studio. Il diametro dei nodi rappresenta il loro grado e il colore (in una scala dal verde al rosso) il livello di inquinamento acustico}
\label{testaccioContrattoColorato}
\end{figure}

\begin{figure}[!ht]
\centering
\includegraphics[width=.8\textwidth]{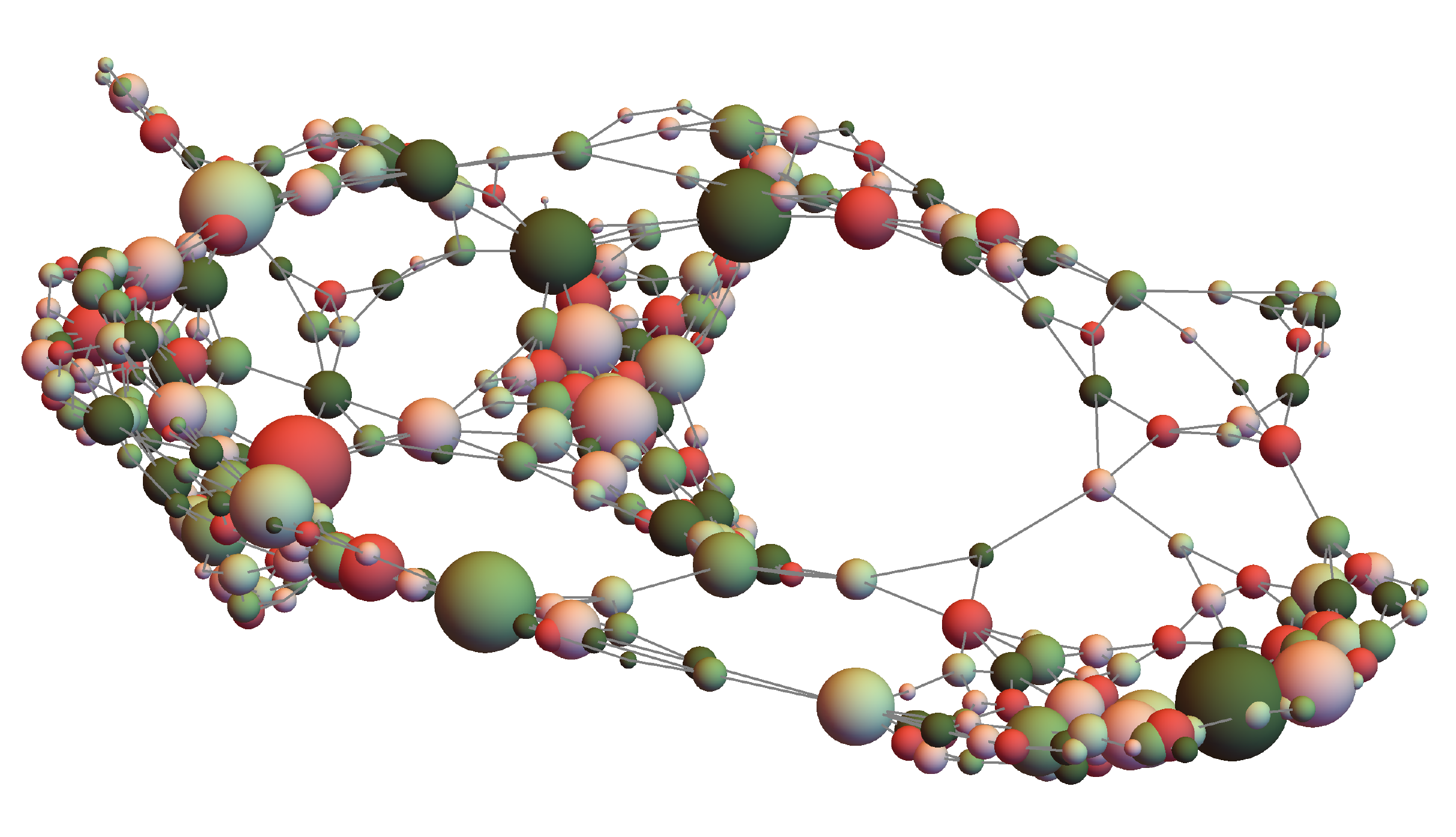}
\caption{Lo stesso grafo di fig.\ref{testaccioContrattoColorato} in 3 dimensioni}
\label{testaccioContrattoColorato3D}
\end{figure}
Allo stesso modo la {\sl global choice} ${\cal C}_g$\cite{Volchenkov2008}\cite{GlossarySS} misura il flusso attraverso uno spazio $i$, cio\`e quante volte un nodo viene attraversato per andare da una parte all'altra della citt\`a, ed \`e definito come: 
\begin{equation}
	{\cal C}_g(i)={\text{\# cammini pi\`u brevi che attraversano } i\over\text{ \# tutti i cammini pi\`u brevi}}
\end{equation}
un nodo ha quindi un valore alto della {\sl global choice} quando ci sono molti {\sl shortest path} che connettono gli spazi che passano attraverso di esso.
In teoria dei grafi ${\cal C}_g(i)$ \`e definita come {\sl betweenness centrality}.
Nel caso acustico ${\cal C}_g(i)$ indica, per esempio, quanto uno spazio con un dato livello di rumore possa attraversato da chi si muove nella citt\`a.

Tutte le grandezze della {\sl space syntax} possono essere generalizzate al caso acustico e rappresentate sul grafo urbano nel piano o nello spazio.
In fig.\ref{testaccioContrattoColorato} e fig.\ref{testaccioContrattoColorato3D} diamo un esempio di questa rappresentazione dove diametro dei nodi del grafo urbano (acustico) del quartiere Testaccio \`e proporzionale al grado, mentre  il colore (dal verde al rosso) indica il livello di inquinamento sonoro.

\section{Conclusioni}
Il metodo proposto e implementato con Mathematica generalizza la {\sl space syntax}.
Il punto essenziale \`e la scelta e la misura di variabili urbanistiche che chiamiamo ``etichette'' associate agli {\sl edge di un grafo}.
In questo lavoro abbiamo considerato solo grandezze scalari, ma le etichette del metodo possono essere generalizzate ad un vettore di variabili urbanistiche.
In questo caso per\`o, per ottenere la contrazione,  deve essere studiato un metodo di confronto tra vettori adiacenti.
La generalizzazione vettoriale del metodo proposto pu\`o risultare anche molto utile per lo studio dell'{\sl hedonic price} \cite{RePEc:ucp:jpolec:v:82:y:1974:i:1:p:34-55}\cite{RePEc:spr:joecth:v:42:y:2010:i:2:p:275-315}.

Nel caso dei grafi acustici dovremmo considerare una dinamica della {\sl space syntax} perch\'e in generale i suoni di una strada mutano in continuazione.
Per ottenere dati realistici \`e necessario calcolarne le propriet\`a statistiche, o la dinamica, attribuendo ai suoni sugli spazi aperti valori distribuiti o la loro dinamica.
Queste analisi possono dare luogo a una statistica sui grafi urbani e a uno studio dell'evoluzione temporale delle loro propriet\`a.

Il metodo proposto permette inoltre di analizzare fenomeni sociali ed economici connessi con la loro collocazione spaziale, come l'uso di servizi, la densit\`a abitativa, i prezzi degli immobili, il grado di scolarizzazione.
Il confronto tra i diversi grafi urbani pu\`o aiutare la comprensione delle relazioni tra questi diversi fenomeni.
Infine il metodo descritto pu\`o servire per realizzare mappe della topologia urbana in relazione a diverse variabili urbanistiche e pu\`o essere di ausilio per la pianificazione di interventi di riqualificazione urbana.\\\\
L'autore vuole ringraziare Laura Velardi e Valerio Palma per la raccolta di dati acustici nelle vie di Testaccio e per le stimolanti osservazioni sui grafi urbani.

\bibliographystyle{plain}
\bibliography{/Users/dautilia-n/Desktop/DTC/WorkInProgress/References/bibliografia_rob.bib} 
\end{document}